# Non-extensive entropy of bosonic Fibonacci oscillators

**Abdullah Algin**[*]

Department of Physics, Eskisehir Osmangazi University, Meselik, 26480-Eskisehir, Turkey

**Abstract**

We discuss possible connections between the thermostatistical properties of a gas of the two-parameter deformed bosonic particles called Fibonacci oscillators and the properties of the Tsallis thermostatistics. In this framework, we particularly focus on a comparison of the non-extensive entropy functions expressed by these two generalized theories. We also show that the thermostatistics of the two-parameter deformed bosons can be studied by the formalism of Fibonacci calculus, which generalizes the recently proposed formalism by Lavagno and Narayana Swamy of $q$-calculus for the one-parameter deformed boson gas. As an application, we briefly summarize some of the recent results on the Bose-Einstein condensation phenomenon for the present two-parameter generalized boson gas.

---

[*] E-mail address: aalgin@ogu.edu.tr

## 1. Introduction

Since the discovery of quantum groups and their associated algebras [1–5] a great deal of effort has been devoted to study their possible applications on many areas of physics such as statistical mechanics. There are two distinct methods in the literature for studying the generalized statistical mechanics. The first method is the use of one or two parameter deformed bosonic and fermionic quantum algebras. The second method is the formalism of Tsallis non-extensive statistical mechanics. In this respect, possible connections between quantum groups and Tsallis non-extensive statistical mechanics have been extensively investigated [6–10].

In the framework of *q*-deformed bosons and similar operators called quons [11], some considerable investigations have been carried out for obtaining a possible violation of the Pauli exclusion principle [12] and also a possible relation to anyonic statistics [13,14]. However, it was recently shown in [15-21] that the high and low temperature thermodynamical properties of the quantum group symmetric bosonic and fermionic oscillator gas models depend on the real deformation parameters. Although, other two-parameter realizations have been studied in the literature [22-24], a complete formalism for the generalization of the thermodynamical and statistical properties of the bosons and fermions coming out by quantum algebraical structures is currently under active investigation.

In this paper, we consider a different generalization of bosonic system, which is called Fibonacci oscillators. They have a spectrum given by a generalized Fibonacci sequence. We focus on the thermostatistical properties of a gas of the commuting Fibonacci oscillators. In particular, we discuss the effect of the deformation parameters $(q_1, q_2)$ on the entropy of the system, and compare than with the results of Tsallis thermostatistics. In this sense, we further study our recent work on the low-temperature thermostatistical behavior of a gas of the commuting Fibonacci oscillators [25], and consider possible connections between the thermostatistical properties of these $(q_1, q_2)$-bosons and the properties of non-extensive quantum statistical mechanics.

<par>Furthermore, we want to show that the thermostatistical properties of these $(q_1, q_2)$-bosons can be studied by using the formalism of Fibonacci calculus, which generalizes the earlier one-parameter deformed formalism of $q$-calculus [26,27]. The results obtained in this way will serve as a two-parameter generalization related to the thermostatistics of earlier $q$-deformed boson gas studies [28-33].

Moreover, another important discussion is on the main reasons for considering two distinct deformation parameters in some physical applications. Although, we reviewed these reasons in many respects in [25,34], some important ones that underline the importance of the usage of Fibonacci oscillators are as follows: Firstly, the Fibonacci oscillators offer a unification of quantum oscillators related to quantum groups [35,36]. They are the most general oscillators having the property of spectrum degeneracy and invariance under the quantum group. In this sense, if the quantum group symmetry is preserved, then the number of deformation parameters in $d$ dimensions should be just two [37]. Secondly, one of the main problems in the theory of quantum groups and algebras is to interpret the physical meaning of the deformation parameters. In this respect, one possible explanation for the deformation parameters $q_1$ and $q_2$ was accomplished by a relativistic quantum mechanical model [35,36]. In such a model, the multi-dimensional Fibonacci oscillator can be interpreted as a relativistic oscillator corresponding to the bound state of two particles with masses $m_1$ and $m_2$. Therefore, the additional parameter $q_2$ has a physical significance so that it can be related to the mass of the second bosonic particle in a two particle relativistic quantum harmonic oscillator bound state. Thirdly, the quantum algebra with two deformation parameters may have more flexibility when dealing with phenomenological applications to the concrete physical models [23,24,38]. Although, any quantum algebra with one or more deformation parameters may be mapped onto the standard single-parameter case [39,40], it has been recently argued that the physical results obtained from a two-parameter deformed oscillator system are not the same [41–43].

Thus, all above considerations concisely give main motivation to consider two distinct deformation parameters, and therefore show the importance and requirement to think the Fibonacci oscillators in physical applications.

The paper is organized as follows: In section 2, we review the basic algebraic and representative properties of the multi-dimensional Fibonacci oscillators. In section 3, we investigate the thermostatistical properties of a gas of the commuting Fibonacci oscillators. In this context, we propose a new method to study the thermostatistics of the $(q_1, q_2)$-bosons through the properties of the Fibonacci calculus. In particular, the entropy of a gas of the commuting Fibonacci oscillators is derived in terms of the real independent deformation parameters $q_1$ and $q_2$. We also discuss possible connections between our results and those obtained with Tsallis thermostatistics. For the sake of completeness, we briefly report some of the recent results derived from an applicationon of these $(q_1, q_2)$-bosons on the Bose-Einstein condensation phenomenon. In the last section, we summarize our results.

## 2. The multi-dimensional bosonic Fibonacci oscillators

In this section, the multi-dimensional two-parameter deformed bosonic Fibonacci oscillator algebra is presented. There are two fundamental types of the multi-dimensional bosonic Fibonacci oscillators: commuting and covariant [35,36]. The two types are related by a transformation and diagonal commutation relations for both types of oscillators are the same. Thus, the algebra generated by the commuting Fibonacci oscillators $a_i$ together with their corresponding creation operators $a_i^*$ is defined by the following deformed commutation relations [35]:

$$[a_i, a_j] = 0, \qquad i, j = 1, 2, \ldots, d,$$

$$[a_i, a_j^*] = 0, \qquad i \neq j,$$

$$\begin{aligned} a_i a_i^* - q_1^2 a_i^* a_i &= q_2^{2N_i} \\ a_i a_i^* - q_2^2 a_i^* a_i &= q_1^{2N_i}, \end{aligned} \qquad (1)$$

$$a_i^* a_i = [N_i], \qquad a_i a_i^* = [N_i + 1],$$

where the spectrum of the deformed boson number operators $[N_i]$ is given by the generalized Fibonacci basic integers as

$$[n_i] = \frac{q_1^{2n_i} - q_2^{2n_i}}{q_1^2 - q_2^2}, \qquad (2)$$

where $q_1$ and $q_2$ are the real positive independent deformation parameters. Hereafter, we will consider $1 < q_1 < \infty$ and $1 < q_2 < \infty$. The generalized Fibonacci basic integer $[n_i]$ is a two-parameter generalization of the following $q$-basic numbers connected with the Arik-Coon $q$-oscillator [44] and the Biedenharn-Macfarlane $q$-oscillator [4,5], respectively:

$$[x] = \frac{q^x - 1}{q - 1}, \qquad [x] = \frac{q^x - q^{-x}}{q - q^{-1}}. \qquad (3)$$

In particular, for the one-dimensional case, the commuting Fibonacci oscillator algebra is defined by the following deformed commutation relations:

$$aa^* - q_1^2 a^* a = q_2^{2N}, \qquad aa^* - q_2^2 a^* a = q_1^{2N},$$

$$[a, a] = 0, \qquad [a^*, a^*] = 0,$$

$$[a, N] = a, \qquad [a^*, N] = -a^*, \qquad (4)$$

where $N$ is the usual boson number operator. However, the spectrum of the deformed boson number operator $a^* a = [N]$ is defined by equation (2). We note that the Fibonacci oscillator algebra has symmetry under the interchange of the deformation parameters $q_1$ and $q_2$. Also, the commuting Fibonacci oscillators are relevant for the construction of coherent states and for constructing unitary quantum Lie algebras [35,36].

On the other hand, the covariant Fibonacci oscillator algebra with the quantum group $SU_{q_1/q_2}(d)$-symmetry is defined by the following deformed commutation relations [35]:

$$c_i c_k = q_1 q_2^{-1} c_k c_i, \qquad i < k,$$

$$c_i c_k^* = q_1 q_2 c_k^* c_i, \qquad i \neq k,$$

$$c_1 c_1^* - q_1^2 c_1^* c_1 = q_2^{2N}, \qquad (5)$$

$$c_k c_k^* - q_1^2 c_k^* c_k = c_{k-1} c_{k-1}^* - q_2^2 c_{k-1}^* c_{k-1}, \qquad k = 2,\ldots d,$$

$$q_1^{2N} = c_d c_d^* - q_2^2 c_d^* c_d,$$

where the total deformed boson number operator for this system is

$$c_1^* c_1 + c_2^* c_2 + \ldots\ldots + c_d^* c_d = [N_1 + \ldots\ldots + N_d] = [N], \qquad (6)$$

whose spectrum is given by the generalized Fibonacci basic integers $[n]$ in equation (2). The deformed bosonic annihilation operators in equation (5) have the representation [37]

$$\begin{aligned}
c_1 &= a \otimes \underbrace{q_2^N \otimes \ldots \otimes q_2^N}_{(d-1)-terms}, \\
c_2 &= q_1^N \otimes a \otimes \underbrace{q_2^N \otimes \ldots \otimes q_2^N}_{(d-2)-terms}, \\
&\ldots\ldots\ldots \\
c_i &= \underbrace{q_1^N \otimes \ldots\ldots q_1^N}_{(i-1)-terms} \otimes a \otimes \underbrace{q_2^N \otimes \ldots \otimes q_2^N}_{(d-i)-terms}, \\
&\ldots\ldots\ldots \\
c_d &= \underbrace{q_1^N \otimes \ldots\ldots q_1^N}_{(d-1)-terms} \otimes a,
\end{aligned} \qquad (7)$$

where the operators $a_i$ and $a_i^*$ satisfy the relations in equations (1). It should be noted that the Fibonacci oscillators have some important limiting cases. They give the multi-dimensional ordinary bosons in the limit $q_1 = q_2 = 1$. The one-parameter deformed bosonic algebra invariant under the quantum group $SU_{q_1}(d)$ can be obtained in the limit $q_2 = 1$ [45]. Also, the multi-dimensional bosonic Newton oscillator algebra invariant under the undeformed group $SU(d)$ can be recovered in the limit $q_1 = q_2 = q^{1/2}$ [46].

Furthermore, the algebra in equation (5) was recently used to investigate the thermostatistics of a gas of such quantum group covariant oscillators in the high and low-temperature limits [17,20,21]. Thus, the covariant Fibonacci oscillators are needed for quantum group invariance and bilinear Hamiltonian with a degenerate spectrum [35,36].

The commuting Fibonacci oscillators in equation (1) possess coherent states. Therefore, we can introduce the basic properties of the Fibonacci difference operator, which plays a central role in the analysis of the Fibonacci calculus [35]. The transformation from Fock observables to the configuration space can formally be accomplished by

$$a_i^* \to z_i, \qquad a_i \to \partial_i^{(q_1,q_2)}, \tag{8}$$

where $\partial_i^{(q_1,q_2)}$ is the Fibonacci difference operator, which can be regarded as a two-parameter generalization of the Jackson derivative (JD) [26,27]. For the sake of simplicity, it can be expressed in the one-dimensional case as [35]

$$\partial^{(q_1,q_2)} f(z) = \frac{f(q_1^2 z) - f(q_2^2 z)}{(q_1^2 - q_2^2)z}, \tag{9}$$

where $f(z)$ is an analytic function, and this equation reduces to the ordinary derivative in the limit $q_1 = q_2 = 1$. Thus, the Fibonacci difference operator generalizes its earlier versions connected with the $q$-basic number definitions given in equation (3). In this sense, the Jackson derivatives (JD) related to the $q$-basic numbers in equation (3) can be obtained from the Fibonacci difference operator in equation (9) by applying the limits $q_2 = 1$ and $q_2 = q_1^{-1}$, respectively.

In the next section, we show that the Fibonacci difference operator plays a central role for studying the thermostatistical properties of a gas of the commuting Fibonacci oscillators. Some important functions of the system such as the entropy can be obtained by using the Fibonacci difference operator in equation (9).

## 3. Thermostatistical properties of the Fibonacci oscillators and connection with Tsallis thermosatatistics

In this section, we first present the thermostatistical properties of the commuting Fibonacci oscillators defined in equations (1) and (2). By focusing on the $(q_1, q_2)$-deformed entropy function of the system, we then investigate possible connections between our results and the results of Tsallis thermostatistics. The system containing the commuting Fibonacci oscillators constitutes essentially a "free" $(q_1, q_2)$-deformed bosonic gas system, since the oscillators do not interact with each other. The reason behind of this consideration is that we do not have both a specific deformed commutation relation between bosonic annihilation (or creation) operators and a quantum group symmetry structure in equation (1). In grand canonical ensemble, the Hamiltonian of such a free $(q_1, q_2)$-deformed bosonic Fibonacci oscillators gas can be expected to have the following form:

$$\hat{H}_{q_1,q_2} = \sum_i (\varepsilon_i - \mu_{q_1,q_2})\hat{N}_i, \tag{10}$$

where $\varepsilon_i$ is the kinetic energy of a particle in the state $i$, $\mu$ is the $(q_1, q_2)$-deformed chemical potential, and $\hat{N}_i$ is the boson number operator relative to $\varepsilon_i$. Similar Hamiltonians were also considered by several other authors [25,28–33,47–64].

The thermal average of an operator is written in the standard form

$$<\hat{O}> = \frac{Tr(\hat{O} e^{-\beta \hat{H}_{q_1,q_2}})}{Z}, \tag{11}$$

where $Z$ is the grand canonical partition function defined as

$$Z = Tr(e^{-\beta \hat{H}_{q_1,q_2}}), \tag{12}$$

where $\beta = 1/kT$, $k$ is the Boltzmann constant, $T$ is the temperature of the system. As in the case of the one-parameter deformed boson gas [28–33], the structure of the density matrix $\rho = e^{-\beta \hat{H}_{q_1,q_2}}$ and the thermal average are undeformed in the present two-parameter boson model. Hence, the structure of the partition function is unchanged. This notion is crucial, since it implicitly amounts to an unmodified structure of the Boltzmann Gibbs entropy,

$$S_{q_1,q_2} = k \ln W_{q_1,q_2}, \tag{13}$$

where $W_{q_1,q_2}$ stands for the number of states of the system corresponding to the set of occupation numbers $\{f_{i,q_1,q_2}\}$. In particular, the number $W_{q_1,q_2}$ is modified in the present two-parameter boson model. It should be noted that this two-parameter generalization is a different deformation from the Tsallis thermostatistics one [6,7], where the structure of the entropy is deformed via the logarithmic function. We will see below the main differences between the entropy of the present $(q_1, q_2)$-boson gas and the Tsallis entropy function in detail.

By following the procedure proposed by [53] for a one-parameter deformed non-interacting boson gas, we derive the mean value of the occupation number $f_{i,q_1,q_2}$. We define the mean occupation number $f_{i,q_1,q_2}$ corresponding to $\hat{N}_i$ by

$$q_1^{2f_{i,q_1,q_2}} = \frac{1}{Z} Tr(e^{-\beta \hat{H}_{q_1,q_2}} q_1^{2\hat{N}_i}), \qquad q_2^{2f_{i,q_1,q_2}} = \frac{1}{Z} Tr(e^{-\beta \hat{H}_{q_1,q_2}} q_2^{2\hat{N}_i}). \tag{14}$$

From equations (1), (2) and (14), we derive

$$[f_{i,q_1,q_2}] = \frac{1}{Z} Tr(e^{-\beta \hat{H}_{q_1,q_2}} [\hat{N}_i]) \equiv \frac{1}{Z} Tr(e^{-\beta \hat{H}_{q_1,q_2}} a_i^* a_i). \tag{15}$$

After applying the cyclic property of the trace [48,49], and using the commuting Fibonacci oscillator algebra in equations (1) and (2), we obtain

$$\frac{[f_{i,q_1,q_2}]}{[f_{i,q_1,q_2} + 1]} = e^{-\beta(\varepsilon_i - \mu_{q_1,q_2})}. \tag{16}$$

From the defination of the generalized Fibonacci basic integer $[n_i]$ in equation (2), we explicitly derive

$$f_{i,q_1,q_2} = \frac{1}{[\ln(q_1^2/q_2^2)]} \ln\left(\frac{z_{q_1,q_2}^{-1} e^{\beta\varepsilon_i} - q_2^2}{z_{q_1,q_2}^{-1} e^{\beta\varepsilon_i} - q_1^2}\right), \tag{17}$$

where $z_{q_1,q_2} = e^{\beta\mu_{q_1,q_2}}$ is the $(q_1,q_2)$-deformed fugacity of the system. This equation may be called as the $(q_1,q_2)$-deformed Bose-Einstein statistical distribution function for a gas of the commuting Fibonacci oscillator. The total number of particles is defined by the constraint $N = \sum_i f_{i,q_1,q_2}$. Obviously, this equation reduces to the standard Bose-Einstein distribution in the limit $q_1 = q_2 = 1$. Also, some of the important limiting cases of the function $f_{i,q_1,q_2}$ should be emphasized here: When we take the limit $q_2 = 1$ and $q_1 = q^{1/2}$, the one-parameter deformed distribution of the Arik-Coon type $q$-bosons [28,30,32] can be obtained. In the limit $q_1 = q^{1/2}$ and $q_2 = q^{-1/2}$, the statistical distribution of the Biedenharn-Macfarlane type $q$-bosons can be recovered [29,31,33,64].

On the other hand, it follows from equations (10) and (12) that the logarithm of the bosonic grand partition function has the form

$$\ln Z = -\sum_i \ln(1 - z_{q_1,q_2} e^{-\beta\varepsilon_i}). \tag{18}$$

This is due to the fact that we choose the Hamiltonian to be a linear function of the boson number operator but it is not linear in $a_i^* a_i$, which can be inferred from equation (1). For this reason, standard thermodynamic relations in the usual form are ruled out as in the case of the one-parameter deformed boson gas [28–33]. For instance, it can be proved that

$$N \neq z\left(\frac{\partial}{\partial z}\right) \ln Z. \tag{19}$$

Here, an important point is to observe that the Fibonacci difference operator defined in equation (9) should be used instead of the ordinary thermodynamics derivative with respect to $z$ as follows:

$$\frac{\partial}{\partial z} \to D_z^{(q_1,q_2)}, \tag{20}$$

where $D_z^{(q_1,q_2)}$ may be called as the modified Fibonacci difference operator:

$$D_z^{(q_1,q_2)} = \frac{(q_1^2 - q_2^2)}{\ln(q_1^2/q_2^2)} \partial_z^{(q_1,q_2)}. \tag{21}$$

Therefore, the total number of particles in the commuting Fibonacci oscillator gas can be expressed as

$$N = z\, D_z^{(q_1,q_2)} \ln Z \equiv \sum_i f_{i,q_1,q_2}, \tag{22}$$

where $f_{i,q_1,q_2}$ is defined by equation (17). Although, the Fibonacci oscillators share most of the nice properties of the $q$-oscillators as is discussed in section 2, the only property which does not hold is the Leibniz rule for the exterior derivation on the Fibonacci-Manin space [35,36]. For this reason, the internal energy $U$ of the commuting Fibonacci oscillators gas can be found by extending the procedure in [28–33] to the present two-parameter case. In this sense, we use the modified Fibonacci difference operator in equation (21) and the ordinary chain rule as

$$U = (-\frac{\partial \ln Z}{\partial \beta}) = \sum_i \frac{\partial y_i}{\partial \beta} D_{y_i}^{(q_1,q_2)} \ln(1 - z_{q_1,q_2} y_i), \tag{23}$$

where $y_i = \exp(-\beta \varepsilon_i)$. This leads to

$$U = \sum_i \varepsilon_i\, f_{i,q_1,q_2}, \tag{24}$$

where $f_{i,q_1,q_2}$ is expressed by equation (17).

All above considerations give the importance of Fibonacci calculus for studying the thermostatistics of the two-parameter deformed bosons to some extent. It should be emphasized that as in the one-parameter deformed boson gas case [28–33], the standard structure of thermodynamics remains exact even in the present two-parameter case.

With the above theoretical motivation in mind, we derive the entropy of the $(q_1, q_2)$-bosons using the thermodynamic expression

$$\frac{S}{k} = \ln Z + \beta U - \beta \mu N, \tag{25}$$

where $U$ and $N$ are defined in equations (22) and (24) by means of the modified Fibonacci difference operator. Hence, using equations (1), (16), and (17), we obtain the entropy of the commuting Fibonacci oscillators gas as follows:

$$S_{q_1,q_2} = k \sum_i \{-n_i \ln[n_i] + (n_i + 1)\ln[n_i + 1] - \ln([n_i + 1] - [n_i])\}, \tag{26}$$

where the generalized Fibonacci basic integer $[n_i]$ is defined in equation (2). The last term in this equation can be approximated in the limit $n \gg 1$ as

$$\ln([n_i + 1] - [n_i]) \approx n_i \ln(q_1 q_2). \tag{27}$$

Therefore, the entropy of our model in equation (26) reduces to the undeformed boson gas entropy in the limit $q_1 = q_2 = 1$ as

$$S_{1,1} = k \sum_i \{-n_i \ln n_i + (n_i + 1)\ln(n_i + 1)\}. \tag{28}$$

In the limit $q_2 = 1$ and $q_1 = q^{1/2}$, the $(q_1, q_2)$-deformed entropy function in equations (26) and (27) gives also the entropy function of the one-parameter deformed boson gas except for the last term which differs only a numerical constant [28,30,32].

By following the procedure proposed by [28–32] for a one-parameter deformed non-interacting boson gas, we can apply the extremization condition to the entropy $S_{q_1,q_2}$ with fixed internal energy and number of particles in order to establish self-consistency of the present two-parameter boson model. Therefore, the extremization condition can be written as

$$\delta(S - \beta U + \beta \mu N) = 0, \tag{29}$$

where $\beta$ and $\beta \mu$ play the role of Lagrange multipliers. This equation can be rewritten as

$$D_{y_i}^{(q_1,q_2)}(S - \beta U + \beta \mu N)\delta y_i = 0. \tag{30}$$

Using equations (21)-(24) and (26), equation (30) leads to the correct two-parameter generalized distribution function as in equation (17) derived from the commuting Fibonacci oscillator algebra.

On the other hand, from equation (13), the entropy of the commuting Fibonacci oscillator gas model is directly proportional to the logarithm of the $(q_1, q_2)$-deformed statistical weight $W_{q_1,q_2}$. We should also note that instead of the ordinary factorial $n!$, the representations of the present two-parameter boson algebra is inversely proportional to the $(q_1, q_2)$-basic factorial $[n]!$ of the generalized Fibonacci basic integer $[n]$. This situation was recently discussed in the construction of the Fock space representations of the Fibonacci oscillator algebra [65]. Hence, we assume that the similar replacement can be attributed to the $(q_1, q_2)$-deformed statistical weight $W_{q_1,q_2}$ as

$$W_{q_1,q_2} = \prod_i \frac{[n_i + g_i - 1]!}{[n_i]! [g_i - 1]!}, \tag{31}$$

where $g_i$ represents the number of subcell levels. Therefore, for the limit $n \gg 1$, we derive an approximation for the $(q_1, q_2)$-basic factorial $[n]!$ as follows:

$$\ln[n]! \approx n \ln[n] - \frac{n^2}{2} \ln(q_1 q_2) - \frac{n}{2}\left[\frac{1}{(q_1^2/q_2^2)^n} + \frac{1}{(q_2^2/q_1^2)^n}\right], \tag{32}$$

where we can neglect the last term as compared with the other two terms in the limit $n \gg 1$. This approximation gives us

$$\ln[n]! \approx n \ln[n] - \frac{n^2}{2} \ln(q_1 q_2), \tag{33}$$

which may be considered as a two-parameter generalized Stirling approximation for the case $(q_1, q_2) > 1$. From equations (13), (31), and (33) the entropy of the commuting Fibonacci oscillators gas can be written as

$$S_{q_1,q_2} = k \sum_i \left\{ n_i \ln \frac{[n_i + g_i]}{[n_i]} + g_i \ln \frac{[n_i + g_i]}{[g_i]} - n_i g_i \ln(q_1 q_2) \right\}, \tag{34}$$

where the generalized Fibonacci basic integer $[n]$ is defined in equation (2). This entropy reduces to the entropy of the undeformed boson gas in the limit $q_1 = q_2 = 1$. Also, when we take the limit $q_2 = 1$ and $q_1 = q^{1/2}$ we obtain the entropy $S_q$ of the one-parameter deformed boson gas except for the last term which differs only a numerical constant [28,30,32].

The entropy in equation (34) has a similar structure as in equation (26) except for the factor $g_i$. However, this factor does not effect the correct form of the $(q_1,q_2)$-deformed Bose-Einstein distribution, since the non-extensive property of the generalized Fibonacci basic integer $[n_i]$ in equation (34) does implicitly guarantee a $g_i$-independent result for the statistical distribution function of the system.

For the sake of completeness, we would like to report briefly the recent results on an application of the present commuting Fibonacci oscillator system to the Bose-Einstein condensation phenomenon [25]. The particle density for the commuting Fibonacci oscillators can be obtained as

$$\frac{1}{\upsilon} = \frac{N}{V} = \frac{4\pi}{h^3}\int_0^\infty p^2 dp \frac{1}{\ln(q_1^2/q_2^2)}\ln\left[\frac{(z_{q_1,q_2}^{-1}e^{\beta p^2/2m} - q_2^2)}{(z_{q_1,q_2}^{-1}e^{\beta p^2/2m} - q_1^2)}\right] + \frac{1}{V}\frac{1}{\ln(q_1^2/q_2^2)}\ln\left[\frac{(1-z_{q_1,q_2}q_2^2)}{(1-z_{q_1,q_2}q_1^2)}\right], \quad (35)$$

which leads to

$$\frac{1}{\upsilon} = \frac{1}{\lambda^3}g_{3/2}(q_1,q_2,z_{q_1,q_2}) + \frac{<n_0>}{V}, \quad (36)$$

where $\lambda = \sqrt{2\pi\hbar^2/mkT}$ is the thermal wavelength, and the average occupation number for the zero-momentum state is

$$<n_0> = \frac{1}{\ln(q_1^2/q_2^2)}\ln\left[\frac{(1-z_{q_1,q_2}q_2^2)}{(1-z_{q_1,q_2}q_1^2)}\right]. \quad (37)$$

The two-parameter generalized Bose-Einstein function $g_n(q_1,q_2,z_{q_1,q_2})$ in equation (36) is defined as

$$g_n(q_1,q_2,z_{q_1,q_2}) = \frac{1}{\Gamma(n)}\int_0^\infty x^{n-1}dx\frac{1}{\ln(q_1^2/q_2^2)}\ln\left[\frac{(z_{q_1,q_2}^{-1}e^x - q_2^2)}{(z_{q_1,q_2}^{-1}e^x - q_1^2)}\right]$$

$$= \frac{1}{\ln(q_1^2/q_2^2)}\left(\sum_{l=1}^\infty \frac{(q_1^2 z_{q_1,q_2})^l}{l^{n+1}} - \sum_{l=1}^\infty \frac{(q_2^2 z_{q_1,q_2})^l}{l^{n+1}}\right), \quad (38)$$

where $x^2 = \beta p^2/2m$. Also, the distribution function $f_{i,q_1,q_2}$ in equation (17) should be non-negative. This results in the following constraints on the $(q_1, q_2)$-deformed fugacity and the chemical potential:

$$z_{q_1,q_2} \leq \begin{cases} q_2^{-4}, & \mu_{q_1,q_2} \leq -4k_B T \ln q_2, & (q_2 > q_1), \\ q_1^{-4}, & \mu_{q_1,q_2} \leq -4k_B T \ln q_1, & (q_2 < q_1), \end{cases} \quad (39)$$

which gives the same constraints on the fugacity and the chemical potential as for the usual boson gas in the limit $q_1 = q_2 = 1$ [66-68]. From equation (36), we should remark that our two-parameter boson model will exhibit the Bose-Einstein condensation when the following condition is satisfied:

$$\frac{\lambda^3}{v} \geq g_{3/2}(q_1, q_2, z_{q_1,q_2}). \quad (40)$$

One can also find a relation between the critical temperature of the commuting Fibonacci oscillator gas and of the undeformed boson gas as follows:

$$\frac{T_c(q_1, q_2)}{T_c(1,1)} = \left(\frac{2.61}{g_{3/2}(q_1, q_2, z_{q_1,q_2})}\right)^{2/3}. \quad (41)$$

In figure 1, we show the plot of equation (41) as a function of the deformation parameters $q_1$ and $q_2$ for the case $(q_1, q_2) \geq 1$. In figure 1, it is interesting to note that when the second deformation parameter $q_2$ increases, the critical temperature of our model increases rather than that of the undeformed boson gas.

On the other hand, the specific heat of the commuting Fibonacci oscillators gas can be obtained from the thermodynamic definition $C_V = (\partial U/\partial T)_{V,N}$. Using the modified Fibonacci difference operator in equation (21), for the limit $T < T_c(q_1, q_2)$, the specific heat of our model can be calculated as

$$\frac{C_V}{Nk} = \frac{15}{4} \frac{(z D_z^{(q_1,q_2)} g_{7/2}(q_1, q_2, z_{q_1,q_2}))}{g_{3/2}(q_1, q_2, z_{q_1,q_2})} \left(\frac{T}{T_c(q_1, q_2)}\right)^{3/2}. \quad (42)$$

For high temperatures, i.e. in the limit $T > T_c(q_1, q_2)$, the specific heat of our model can be approximated as

$$\frac{C_V}{Nk} \approx \frac{3}{2}\frac{(q_1^2 - q_2^2)}{\ln(q_1^2/q_2^2)} + \frac{3}{4.2^{9/2}}[2](5[2]+4)\, g_{3/2}(q_1,q_2,z_{q_1,q_2})\left(\frac{T_c(q_1,q_2)}{T}\right)^{3/2}, \quad (43)$$

where the generalized Fibonacci difference operator $[n]$ is defined in equation (2). In figure 2, we plot the specific heat $C_V/Nk$ as a function of $T/T_c(q_1,q_2)$ for values of the deformation parameters $q_1$ and $q_2$ for the case $(q_1,q_2)\geq 1$. Thus, the specific heat of the commuting Fibonacci oscillators gas shows a discontinuity at the critical temperature as shown in figure 2. According to figure 2, when the second deformation parameter $q_2$ increases, the discontinuity in the specific heat of the system also increases. Obviously, in the limit $q_1 = q_2 = 1$, such a discontinuity disappears as in the case of the undeformed boson gas.

From equation (25), we also obtain the entropy of the commuting Fibonacci oscillators gas in the high temperature limit as follows:

$$\frac{S_{q_1,q_2}}{V} = \frac{k}{\lambda^3}\left[\frac{5}{2} g_{5/2}(q_1,q_2,z_{q_1,q_2}) - g_{3/2}(q_1,q_2,z_{q_1,q_2})\ln z_{q_1,q_2}\right]. \quad (44)$$

For low temperatures, the entropy of our model becomes

$$\frac{S_{q_1,q_2}}{V} = \frac{5}{2}\frac{k}{\lambda^3} g_{5/2}(q_1,q_2,z_{q_1,q_2}). \quad (45)$$

From equations (44), (45), and (39), the jump of the entropy at $T = T_c(q_1,q_2)$ can be obtained as

$$\frac{\Delta S_{q_1,q_2}}{V} = \frac{4k}{\lambda^3} g_{3/2}(q_1,q_2,z_{q_1,q_2})\ln(q_1 q_2). \quad (46)$$

The entropy of the commuting Fibonacci oscillators gas gives the same results as the entropy of an undeformed boson gas in the limit $q_1 = q_2 = 1$ [66-68]. We observe from equation (46) that for $(q_1,q_2)>1$ the entropy values of the commuting Fibonacci oscillators gas at the critical point is different than those of an undeformed boson gas. The jump of the entropy of the commuting Fibonacci oscillators gas at the critical point increases with the values of the deformation parameters $q_1$ and $q_2$. Moreover, the results in equations (42) and (45) are compatible with the third law of thermodynamics in the limit $T \to 0$.

We can further study other results for the classical limit $z_{q_1,q_2} = e^{\beta \mu_{q_1,q_2}} \ll 1$. For instance, in this limit, the $(q_1, q_2)$-deformed Bose-Einstein functions $g_n(q_1, q_2, z_{q_1,q_2})$ in equation (38) reduce to

$$g_n(q_1, q_2, z_{q_1,q_2}) = \frac{(q_1^2 - q_2^2)}{\ln(q_1^2/q_2^2)} z_{q_1,q_2}. \qquad (47)$$

Also, from equation (43), the specific heat of the commuting Fibonacci oscillators gas reduces to the following expression in the classical limit:

$$\frac{C_V}{Nk} \approx \frac{3}{2} \frac{(q_1^2 - q_2^2)}{\ln(q_1^2/q_2^2)}, \qquad (48)$$

which shows that the two-parameter deformation remains exact even in the classical limit. This feature also exists in the case of Tsallis thermostatistics.

With the light of above discussion, we should remark that the results in equations (10)-(48) are not only different from the results of the one–parameter deformed boson model studied in [28–33], but also do they generalize the results to the case with two deformation parameters via the Fibonacci oscillators.

Now, we would like to discuss possible connections between the entropy $S_{q_1,q_2}$ of our model given in equation (26) and the Tsallis entropy $S_{\tilde{q}}^T$. In order to summarize the Tsallis thermostatistics, let us recall the axioms on which the formalism is based [6,7]: (i) The thermal average of an operator is given by

$$<\hat{O}>_T = \frac{Tr(\rho_{\tilde{q}}^T \hat{O})}{Tr(\rho_{\tilde{q}}^T)}, \qquad (49)$$

where $\rho_{\tilde{q}}^T$ is the Tsallis distribution function. (ii) The entropy of the system is defined to be

$$S_{\tilde{q}}^T = k \ln_{\tilde{q}}^T W \equiv k \frac{W^{1-\tilde{q}} - 1}{1 - \tilde{q}}. \qquad (50)$$

For both equations in (49) and (50), the standard expressions can be recovered in the limiting case of a new parameter $\tilde{q}$ called entropic index as $\tilde{q}=1$. Therefore, we should emphasize that the Tsallis entropy $S_{\tilde{q}}^{T}$ is based on the deformation of the logaritmic function of the Boltzmann entropy $S=k\ln W$.

However, the $(q_1,q_2)$-deformed entropy $S_{q_1,q_2}$ in equation (26) is based on the deformation of the statistical weight via the modification

$$W \to W_{q_1,q_2}, \quad (51)$$

where $W_{q_1,q_2}$ is the two-parameter deformed statistical weight expressed by means of the generalized Fibonacci basic integer $[n]$ in equation (31). As a result, our two-parameter deformed entropy for the commuting Fibonacci oscillators gas and the Tsallis entropy basically show different generalized theories. Both of them are non-extensive but, as discussed above, they imply different consequences of thermostatistics.

Although, the Tsallis thermostatistics does not embody any quantum group and its associated algebraic structure, it shares with quantum groups the mathematical formalism of $q$-calculus, which can be recovered from the present Fibonacci calculus in the limit $q_2=1$, $q_1=q^{1/2}$. We should outline that both the Tsallis entropy and the generalized Fibonacci basic integer $[n]$ in the $(q_1,q_2)$-entropy of our model has the non-additivity property. We should also add that the $(q_1,q_2)$-entropy of our model in equation (26) does not reduce to the standard Boltzmann-Gibbs entropy $S=-k\sum_{i} p_i \ln p_i$ in the classical limit, except for the limit $q_1=q_2=1$.

Now, we would like to continue to consider possible connections between the bosonic Fibonacci oscillators and the Tsallis thermostatistics in much more details. In this context, a detailed discussion on possible connections between the two generalized theories can be categorized into two groups as follows: (i) The first one is related to the properties of the commuting Fibonacci oscillators given in equations (1) and (2). The connection between the Tsallis entropy and the Jackson derivative (JD) is well-known in the literature [6–10,70–76]. Therefore, it is natural to pursue a parallel way for obtaining a connection between the two-parameter generalized entropy of our model and the Tsallis entropy. This can be accomplished through the properties of the Fibonacci calculus discussed in section 2.

Recently, Abe has proposed the following connection between the Tsallis entropy $S_{\tilde{q}}^T$ and the JD [8,71]:

$$S_{\tilde{q}}^T = -k\, \partial_x^{(\tilde{q})} \sum_i P_i^x \bigg|_{x=1} = -k \sum_i \frac{P_i^{\tilde{q}} - P_i}{\tilde{q} - 1}, \qquad (52)$$

where $\partial_x^{(\tilde{q})}$ is the JD defined as

$$\partial_x^{(\tilde{q})} f(x) = \frac{f(\tilde{q}x) - f(x)}{x(\tilde{q} - 1)}, \qquad (53)$$

which can be recovered from the Fibonacci difference operator in equation (9) in the limit $q_2 = 1$ and $q_1 = \tilde{q}^{1/2}$. Also, the JD is intimately connected with the $q$-basic number definitions given in equation (3) [77]. By means of the Fibonacci difference operator $\partial^{(q_1, q_2)}$ defined in equation (9), a two-parameter generalized entropy can be deduced as

$$S_{q_1, q_2} = -k\, \partial_z^{(q_1, q_2)} \sum_i P_i^z \bigg|_{z=1} = -k \sum_i \frac{P_i^{q_1^2} - P_i^{q_2^2}}{q_1^2 - q_2^2}. \qquad (54)$$

We note that this two-parameter deformed entropy has symmetry under the interchange of the deformation parameters $q_1$ and $q_2$. The expressions for the entropy in equations (26) and (54) satisfy the non-additivity property formally similar to the generalized Fibonacci basic integer [n] in equation (2) such as [35]

$$[n + m] = [m][n + 1] + [m + 1][n] - (q_1^2 + q_2^2)[m][n], \qquad (55)$$

which, in the limit $q_1 = q_2 = 1$, reduces to the sum of the ordinary numbers *n, m*, and all such relations reduce to the undeformed boson gas results.

The two-parameter generalized entropy function in equation (54) may be expressed in terms of the Tsallis entropy as

$$S_{q_1,q_2} = \frac{(q_1^2-1)S^T_{q_1^2} - (q_2^2-1)S^T_{q_2^2}}{(q_1^2 - q_2^2)}. \tag{56}$$

This expression has some important limiting cases: In the limit $q_2 = 1$ and $q_1 = \tilde{q}^{1/2}$, the Tsallis entropy function in equation (52) can be recovered [6,7]. When we take the limit $q_2 = q^{-1/2}$ and $q_1 = q^{1/2}$, the symmetric Tsallis entropy with the $q \leftrightarrow q^{-1}$ invariance can be obtained [8,71,72,75]. In the limit $q_2 = (q')^{1/2}$ and $q_1 = q^{1/2}$, the $(q,q')$-entropy proposed by [72] can also be obtained. Obviously, the limit $q_1 = q_2 = 1$ recovers the standard Boltzmann-Gibbs entropy.

However, we should emphasize that our new entropy $S_{q_1,q_2}$ in equation (54) is originally different generalization from the entropy $S_{q,q'}$ studied in [72], which was based on the one-dimensional $(p,q)$-oscillator structure proposed by Chakrabarti and Jagannathan [22]. Since, our analysis is based on the Fibonacci calculus through the properties of the multi-dimensional bosonic Fibonacci oscillators as discussed in section 2. Therefore, we should remark that the multi-dimensional bosonic Fibonacci oscillators give a general framework and a unification of the non-extensive entropies related to quantum groups.

Moreover, equations (54)-(56) give a general relationships among different $q$-oscillator structures in connection with the Tsallis thermostatistics. The relations in equations (54)-(56) are also very important in the sense that they could establish a basic recipe to study deeply other possible connections between the multi-dimensional Fibonacci oscillators and the Tsallis thermostatistics. In a general framework, we should also observe that both the $(q_1,q_2)$-boson thermostatistics and the Tsallis thermostatistics are connected with the Fibonacci difference operator and the Fibonacci calculus. (ii) The second one is related to the peperties of the covariant Fibonacci oscillators given in equations (5)-(7). Recently, it has been shown in [78,79] that the non-extensivity of classical set theory expressed by

$$m(A \cup B) = m(A) + m(B) - m(A \cap B), \tag{57}$$

turns out to have a *q*-oscillator structure. This observation was also related to unitary quantum groups such as the $U_q(d)$. Using this non-extensivity property, a *q*-distribution was obtained [78,79]. Therefore, as the bosonic Fibonacci oscillators offer a unification of oscillators related to quantum groups, the covariant Fibonacci oscillator algebra could play new roles, and could give some more general results in connection with the non-extensive physics. Furthermore, by investigating a possible form of the two-parameter generalized probability distribution, one may further establish a possible link between the covariant Fibonacci oscillators and the Tsallis thermostatistics.

Although the role of quantum group invariance in non-extensive quantum statistical mechanics has been studied by means of the one-parameter quantum group structures [80,81], it has been argued in [17–25,34–38,43,65] that the two-parameter quantum group boson gas studies could give different consequences of thermostatistics. Following this motivation, one could further analyze this relevant point as follows: One can construct the two-parameter quantum group covariant density matrix by using the covariant Fibonacci oscillator algebra generators in equations (5)-(7). Then, one would hopefully investigate possible consequences of imposing a two-parameter density matrix to the formalism of the Tsallis thermostatistics.

All above considerations show that both the quantum algebraic and the thermostatistical properties of the bosonic Fibonacci oscillators deserve more work to deduce some closer connections with the formalism of the Tsallis thermostatistics. In particular, if one consider the relevance of taking two distinct deformation parameters in physical applications discussed in section 1, an analysis on the non-extensive thermostatistics with the Fibonacci oscillators could then give new interesting results in the field of quantum statistical mechanics.

## 4. Summary

In this study, we discussed possible connections between the thermostatistics of two-parameter generalized bosonic Fibonacci oscillators and the Tsallis thermostatistics. In this framework, we have shown that the thermostatistics of the $(q_1, q_2)$-bosonic particles can be studied using the formalism of Fibonacci calculus. This consideration also allows us to generalize the recently proposed formalism of $q$-calculus for the one-parameter deformed boson gas [28-33]. The form of the $(q_1, q_2)$-deformed entropy of our model in equation (26) provides the necessary information on a relation between the Fibonacci oscillator algebra and the quantum statistical behavior. In addition, in the classical limit, the $(q_1, q_2)$-deformed entropy and the specific heat of the system remains deformed as in the case of the Tsallis thermostatistics.

Moreover, starting with the $(q_1, q_2)$-deformed Bose-Einstein statistical distribution function and the modified Fibonacci difference operator, we briefly reported the conditions under which the Bose-Einstein condensation would ocur in the present two-parameter generalized boson model. The specific heat of this model has a $\lambda$-point transition behavior which is not exhibited by an undeformed boson gas.

Although, there is no evidence as to why quantum mechanics of bosons and fermions should be different than that of the standard (undeformed) variety, there are some phenomenological studies giving reasons for the use of quantum algebras as outlined in the introduction section. In this sense, one might well view $(q_1, q_2)$-deformation as a phenomenological means of introducing extra parameters to account for non-linearity in the system under consideration. Therefore, in order to support this point of view, we can examplify a realistic model as follows: For values of the deformation parameters $q_1 \approx 1.06$ and $q_2 \approx 1.58$, the behavior of the specific heat of our model is qualitatively good agreement with the experimental data on the gap in the specific heat of a dilute gas of rubidium atoms [69]. Hence, models covering two-parameter generalized bosonic Fibonacci oscillators could give new implications in understanding of non-linear behaviors of realistic systems especially in the condensed matter physics.

**Acknowledgments**

The author thanks the referee for useful comments. This work is supported by the Scientific and Technological Research Council of Turkey (TUBITAK) under the project number 106T593.

**Figure Captions**

**Figure 1.** The ratio $T_c(q_1,q_2)/T_c(1,1)$ of the $(q_1,q_2)$-deformed critical temperature $T_c(q_1,q_2)$ and the undeformed $T_c(1,1)$ as a function of the deformation parameters $(q_1,q_2)$ for the case $(q_1,q_2) \geq 1$.

**Figure 2.** The specific heat $C_V/Nk$ as a function of $T/T_c(q_1,q_2)$ for various values of the deformation parameters $q_2$ and $q_1 \geq 1$.

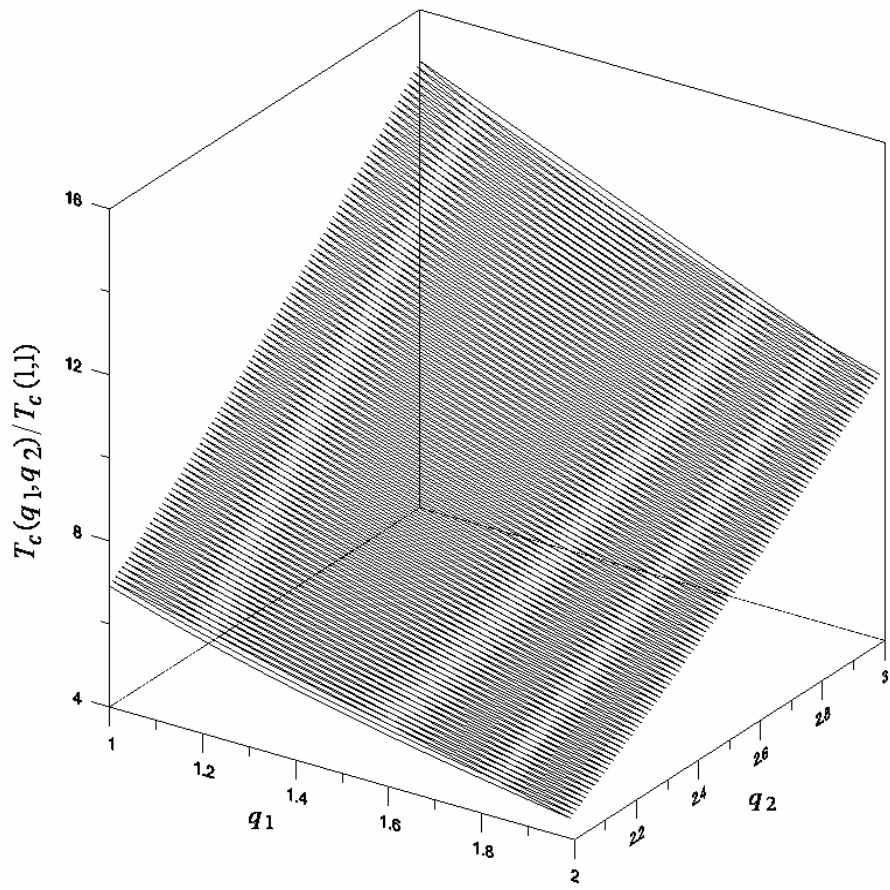

**Figure 1.**

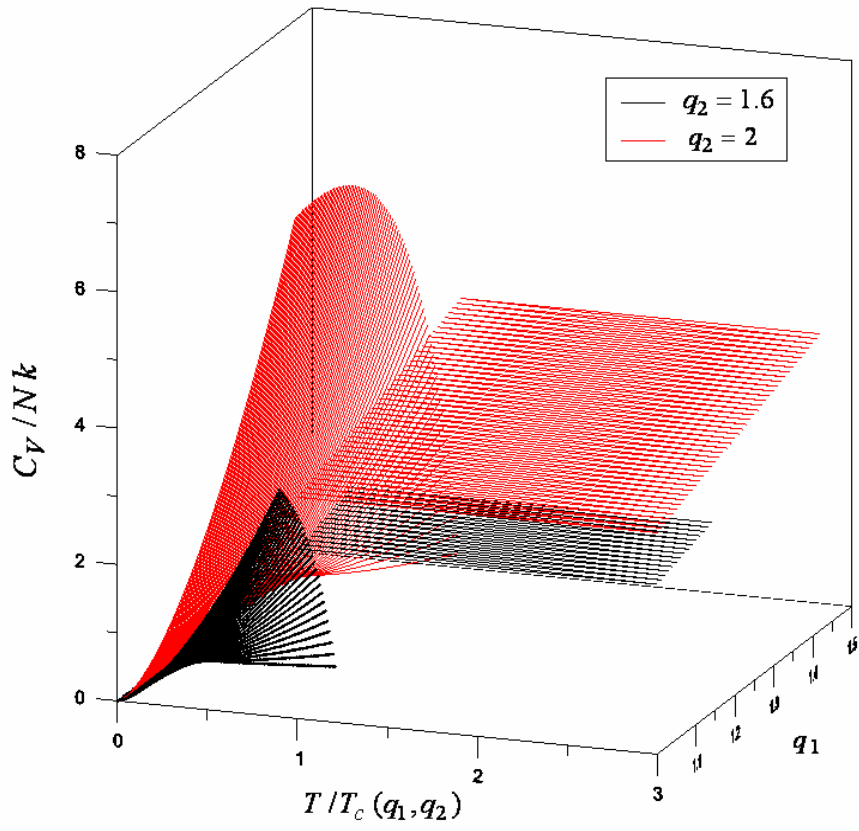

**Figure 2.**